\documentclass[aps,prb,twocolumn,superscriptaddressgroupaddress,floatfix]{revtex4-2}
\usepackage{amssymb}
\usepackage{physics}
\usepackage{graphicx}
\usepackage{times}
\usepackage{amsmath}
\usepackage{amsthm}
\usepackage{amsfonts}
\usepackage[T1]{fontenc}
\usepackage[latin9]{inputenc}
\usepackage{array}
\usepackage{multirow}
\usepackage{color}
\usepackage{bm}
\usepackage{color}

\usepackage{bbm}
\usepackage{hyperref}
\usepackage{babel}
\usepackage{float}
\allowdisplaybreaks[4]
\begin{document}
	\title{Localization spectrum of a bath-coupled generalized Aubry-Andr\'e model in the presence of interactions}
	\author{Yi-Ting Tu}
	\author{DinhDuy Vu}
	\author{Sankar Das Sarma}
	\affiliation{Condensed Matter Theory Center and Joint Quantum Institute, Department of Physics, University of Maryland, College Park, Maryland 20742, USA}
	\begin{abstract}
          A generalization of the Aubry-Andr\'e model, the non-interacting GPD model introduced in [S.\ Ganeshan {\it et al.}, \href{https://doi.org/10.1103/PhysRevLett.114.146601}{Phys.\ Rev.\ Lett.\ {\bf 114}, 146601 (2015)}], is known analytically to possess a mobility edge, allowing both extended and localized eigenstates to coexist. This mobility edge has been hypothesized to survive in closed many-body interacting systems, giving rise to a new non-ergodic metallic phase. In this work, coupling the interacting GPD model to a thermal bath, we provide direct numerical evidence for multiple qualitative behaviors in the parameter space of disorder strength and energy level. In particular, we look at the bath-induced saturation of entanglement entropy to classify three behaviors: thermalized, non-ergodic extended, and localized. We also extract the localization length in the localized phase using the long-time dynamics of the entanglement entropy and the spin imbalance. Our work demonstrates the rich localization landscape of generalized Aubry-Andr\'e models containing mobility edges in contrast to the simple Aubry-Andr\'e model with no mobility edge. 
	\end{abstract}

	\maketitle
	
	\section{Introduction}
        Systems with many-body localization (MBL) retain some information of the initial state in the long-time limit, evading the eventual thermalization postulated by the Eigenstate Thermalization Hypothesis \cite{Altman2015_Review, Nandkishore2015, Abanin2019, Gopalakrishnan2020_Review}. Many-body localization was first realized on one-dimensional systems with random disorder \cite{Basko2006, Oganesyan2007_PRB, Znidaric2008, Pal2010, Devakul2015, Imbrie2016, Imbrie2016b} as a natural extension of the non-interacting Anderson localization \cite{Anderson1958, Abrahams1979} into the interacting regime. Some analytical advancement are obtained by study randomized Bethe lattice \cite{Basko2006,DeLuca2014,Aizenman2011}. However, at the single-particle level, localization can also emerge in deterministic quasi-periodic Hamiltonians with the earliest example of the Aubry-Andr\'e (AA) model \cite{Aubry1980, Harper1955}. Superficially, both random and quasi-periodic potentials break the translational symmetry, but their localization properties are qualitatively distinct, starting already at the single-particle level. In particular, while the Anderson model in one dimension is always localized, the quasi-periodic AA has a critical disorder strength for inducing localization characterized by a self-duality. The self-dual point of quasi-periodic models can be decorated with an eigenstate energy dependence \cite{Soukoulis1982, Boers2007, Biddle2011, Bodyfelt2014, Ganeshan2015, Liu2015, Gopalakrishnan2017, Li2020, Wang2020, Wang2023, Vu2023}, giving rise to the single-particle mobility edge - the energy level separating localized and extended single-particle eigenstates. Since single-particle quasi-periodic localization has its own MBL generalization \cite{Vidal2002, Iyer2013, Mastropietro2015, Khemani2017_PRL, Xu2019_PRR, Vu2022} in the presence of interactions, one natural question is whether the coexistence of both localized and thermalized eigenstates survives interaction \cite{Modak2015, Li2015a, Nag2017, Hsu2018_PRL, An2021, Huang2023}, or equivalently, the fate of the single-particle mobility edge in the corresponding interacting quasiperiodic Hamiltonian. However, the study of non-equilibrium physics even in one-dimensional interacting systems of interest in the current work is mostly limited to small-size numerical simulation, and therefore making a conclusive analytic statement is not possible at this stage.  One possible scenario supported by numerics is that the single-particle mobility edge is destroyed or pushed to the low and high-energy tails of the spectrum \cite{Huang2023}.
	
	One notable example of the class of quasiperiodic systems with mobility edges is the generalization of AA potential (see Eq.~\ref{eq:pot}) introduced in Ref.\ \cite{Ganeshan2015} that produces an analytical single-particle mobility edge.
	We refer to this particular generalized AA model as the GPD model, emphasizing the fact that this is a specific generalization, and other generalized AA models with mobility edges also exist (see, e.g.,~\cite{Soukoulis1982,Boers2007,Biddle2011,Bodyfelt2014}).
	Surprisingly, unlike the general hypothesis we mentioned earlier, existing numerical evidence \cite{Xu2019_PRR,Vu2022,Modak2015,Li2015a,Nag2017,Hsu2018_PRL} for the GPD model suggests that the single-particle mobility edge survives interactions and the interacting system manifests something like a many body mobility edge separating the many body spectrum of the interacting GPD model. 
In Ref.~\cite{Li2015a,Li2016}, the authors studied the entanglement entropy scaling law of each eigenstate and found an area-volume law transition with respect to the eigenenergy, marking a localized-extended crossover in the interacting GPD system. Similarly, the eigenstate fluctuation - variation among the expectation value of a local operator on neighboring eigenstates - vanishes after some energy level upon entering the ergodic regime~\cite{Li2015a,Li2016}. Notably, the two characteristic energies do not match, suggesting three phases on the spectrum: MBL, non-ergodic extended, and ETH (ergodic extended). While the chaotic extended and integrable localized phases are usually thought of as the only two regimes, the non-ergodic extended phase has been demonstrated on the disordered Bethe lattice \cite{DeLuca2014}. Reference~\cite{Hsu2018_PRL} and \cite{YiTing_ML} extended the idea by applying machine learning classification on the eigenstate entanglement spectrum. They found that the three-output classification scheme is optimal, reasserting the existence of an intermediate phase which is neither MBL nor ETH in the interacting GPD spectrum. In the current work, we study the localization spectrum of the many-body GPD from a new perspective - the stability of the interacting system against a bath-induced avalanche. In our approach, the system is coupled to a bath, which sets it apart from the previous papers that study isolated systems. Additionally, we look at the dynamic evolution of the interacting system rather than static quantities as was done before.
	
The bath-coupling approach was first used to answer the question whether MBL survives in the thermodynamic limit \cite{DeRoeck2017, Thiery2018, Crowley2020, Morningstar2022, Sels2022, Tu2023, Peacock2023}. The intuition is as follows. In an infinitely long chain with random disorder, rare large thermal regions may exist which keep expanding until the entire system is delocalized, suppressing MBL. References~\cite{Morningstar2022, Sels2022} coupled the spin chain to an ideal infinite temperature bath to mimic such a rare thermal inclusion and studied the scaling of the thermalization rate with respect to the chain length, with the specific question being the existence of a possible avalanche instability destroying MBL. For quasi-periodic systems, due to the deterministic nature of the system, the avalanche is unlikely to occur spontaneously, making quasi-periodic MBL more stable than the random disorder case in large-size systems \cite{Tu2023}. Nevertheless, one can still think of the bath-induced avalanche as a dynamic probing tool, in the same manner as in the quenched dynamics experiments \cite{Luschen2018, Schreiber2015, Kohlert2019}. Indeed, artificially implanted thermal seeds have been implemented experimentally to directly study the possibility of the avalanche~\cite{Leonard2023}. 
We use this same technique of introducing a bath-coupling to study the stability (or not) of the mobility edge in the interacting GPD model in the current work.
	
	In this paper, we numerically simulate an experiment in which a spin chain under the quasi-periodic GPD potential (system) is coupled to another spin chain without the disorder (bath). The system's initial state is chosen to be an eigenstate of the respective Hamiltonian so that without the bath, the initial state is strictly invariant and survives forever. Upon being coupled to the bath, the system eventually thermalizes, characterized by the saturation of the entanglement entropy. However, depending on the localization property of the initial eigenstate, this process can be either fast, slow, or decelerated (fast at a short time but slower at a longer time), which allows us to categorize the initial state into ETH, MBL, and the exotic non-ergodic metallic phase.
Our goal is to find (or not) a clear behavior in the system coupled to the bath which is manifestly intermediate between the well-known ETH and MBL behaviors.  This does not happen either in the random disorder Anderson model or the quasiperiodic AA model, where numerics only find  MBL or ETH phases in the interacting system depending on the disorder strength.

It is possible that MBL itself is a finite-time finite-system transient, which disappears in the thermodynamic limit, and if so, the same would happen to all the associated physics including that in the current work.  While the fate of MBL (and all associated generic quantum phenomena in interacting disordered systems) remains an important open question of principle well beyond the scope of the current work, it is undeniable that MBL phenomena manifest themselves in finite size experimental atomic and qubit systems in apparent agreement with the theoretical work. Therefore, in the same spirit as Refs.~\cite{Li2015a, Hsu2018_PRL,Li2016}, our ``phase'' refers to a finite-size finite-time manifestation, and we refrain from commenting on the thermodynamic limit, which is inaccessible to any current simulations (and which remains open even for the crucial question of the existence or not of MBL itself). We also study the situation where the system is initialized in the N\'eel state, which is a superposition of almost all eigenstates. In the non-interacting systems, AA is more stable than GPD against bath-induced thermalization since AA has no mobility edges. Surprisingly, the situation reverses in the presence of interaction, now the AA model being more susceptible to the bath. Again, this result is consistent with recent findings on the early localization of the closed GPD model, which establishes that interaction actually enhances MBL in GPD \cite{LiGPD}.
	
	The rest of this paper is organized as follows. In Sec.~\ref{setup}, we provide details of the setup. We analyze the simulation results on entanglement growth in Sec.~\ref{threephases} using a log-log ansatz, sorting each set of parameters into one of the behaviors (ETH or MBL or non-ergodic extended). In section~\ref{llength}, we focus on the intermediate and strong disorder regimes and use a linear-log ansatz to extract the localization length from entanglement entropy and spin imbalance dynamics. A prominent energy dependence shown in the interacting GPD compared to the AA model reiterates the phase classification of Sec.~\ref{threephases}. We conclude and discuss the outlook in Sec.~\ref{conclude}. 
	
	\section{Setup and methods}\label{setup}
	
	\begin{figure*}
		\includegraphics[trim=0 0 0 0, clip, scale=0.62]{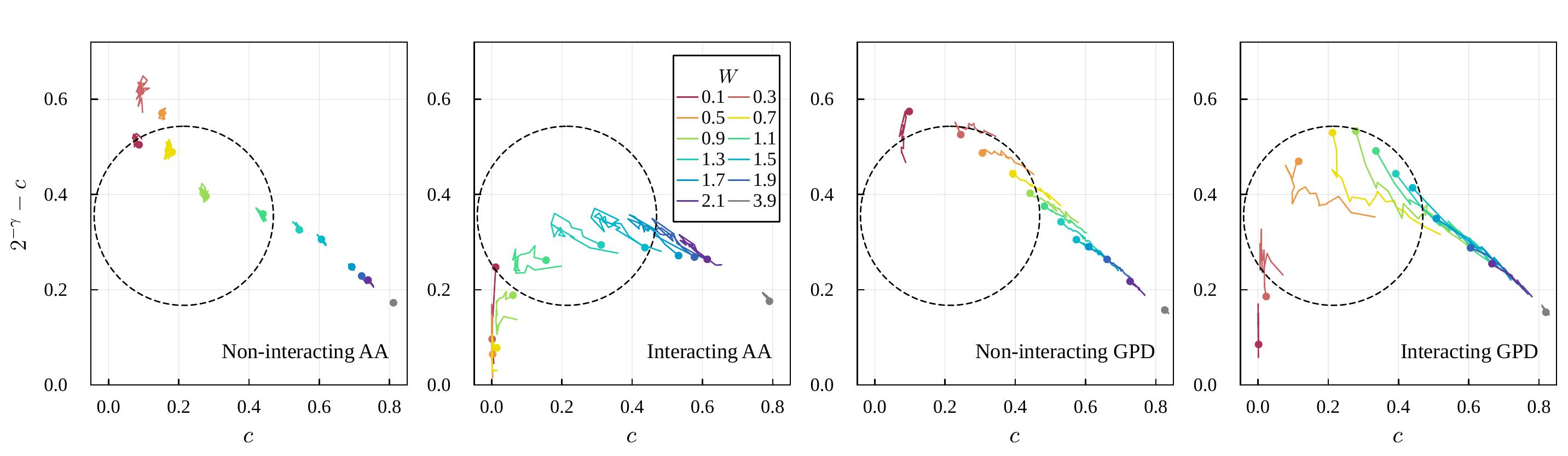}
		\caption{
                  The behavior of the four systems in the $(c,2^{-\gamma}-c)$ feature space. The lower-left, upper-left, and lower-right correspond to the ETH, the non-ergodic extended, and the localized behavior, respectively.
			The colored dots indicate the behavior at the highest part of the energy spectrum, and the ``tails'' of the dots indicate the behavior across the energy spectrums, where the end of the tails correspond to the lowerest parts.
			Dashed circles indicate where the three behaviors cross over.
			Note the effect of the mobility edge in non-interacting GPD (the upper part of the spectrum is more localized).
			For the interacting GPD where the behavior is not yet fully understood, we can see that the upper part of the spectrum goes to the non-ergodic extended phase in intermediate disorder, suggesting a non-ergodic metallic phase.
		}
		\label{fig:circle}
	\end{figure*}
	
	We focus on a $1/2-$spin chain described by the Hamiltonian
	\begin{equation}\label{eq:sys}
		H_{s} = \sum_{j=1}^{L_{s}-1} \left( S_j^x S_{j+1}^x + S_j^y S_{j+1}^y + V S_j^z S_{j+1}^z \right) + W\sum_{j=1}^{L_s} h_j S_j^z,
	\end{equation}
        where $S^{x,y,z}_j$ is the spin-$1/2$ operators at site $j$, $V$ is the interaction strength, and $W$ is the disorder strength. This paper focuses on $V=0$ (non-interacting) and $V=1$ (interacting). The on-site potential is defined by the GPD quasiperiodicity~\cite{Ganeshan2015}
	\begin{equation}\label{eq:pot}
		h_j = \frac{\cos(2\pi\varphi j + \phi)}{1-\alpha\cos(2\pi\varphi j + \phi)} + \text{const.},
	\end{equation}
        where $\varphi=\frac{1+\sqrt{5}}{2}$ is the golden mean, $\phi$ is an initial phase, which is randomly sampled, and the system size is $L_s=12$. The constant is chosen such that $h_j$ sum to zero. In this paper, we compare two representative cases of Eq.~\ref{eq:pot}, one without a mobility edge (AA) and one with a mobility edge (GPD): $\alpha=0$ and $\alpha=-0.8$. In summary, we study four cases: non-interacting AA, interacting AA, non-interacting GPD, and interacting GPD. The behaviors of the non-interacting counterparts are well established, which we use to benchmark the more elusive interacting systems. In particular, the entire non-interacting AA spectrum is localized (extended) for $W>1(<1)$. On the other hand, the single-particle GPD model has an intermediate phase with a mobility edge where the spectrum accommodates localized and extended eigenstates for $0.1\lesssim W \lesssim 2.0$ \cite{LiGPD}.
	
	To simulate the thermal bath, we consider another non-disordered Heisenberg spin chain
	\begin{equation}\label{eq:bath}
		H_b = \sum_{j=1-L_b}^{-1} \left( S_{j}^x S_{j+1}^x + S_{j}^y S_{j+1}^y + S_{j}^z S_{j+1}^z \right),
	\end{equation}
        where $L_b=12$ is the same as the size of the system.
          Although the Heisenberg spin chain is Bethe-ansatz-integrable and technically non-ergodic. We numerically test that adding an integrability-breaking next-nearest-neighbor hopping term (with coefficient 0.2 similar to Ref.~\cite{Goihl2019}) yields negligible effect. Therefore, it is safe to consider the Heisenberg spin chain as a thermal bath in our setup.

      We couple the bath to the system. Here, the system-bath coupling is modeled by a repeating sequence of sharp pulses so that the dynamics of every cycle $\tau=10$ is described by first independent evolutions of the bath and system under their respective Hamiltonians, that is, by $\exp(-i\tau H_s)\exp(-i\tau H_b)$, and then the coupling unitary
	\begin{equation}
		U_\text{sb} = \exp\left[-i\tau (S_0^x S_1^x + S_0^y S_1^y + S_0^z S_1^z)\right]
	\end{equation}
	that entangles the last spin of the bath and the first spin of the system.
        We note that the system-bath coupling is implemented entirely through the unitary $U_{sb}$, and not through any  ad hoc terms in the Hamiltonian.
        This setup allows for all the evolution unitaries $U_b$, $U_s$, and $U_{sb}$ to be computed exactly using exact diagonalization in advance, allowing us to evolve the composite system to a sufficiently long time. This numerical method is equivalent to the TEBD-type algorithm used in~\cite{Peacock2023}.
        We implement this algorithm by the ITensor Julia package~\cite{ITensor} with cutoff $10^{-10}$.

	The energy-resolved initialization is a significant difference between our setup from other MBL dynamical probes. While the initial state is usually the N\'eel state, which is the superposition of an extensive number of eigenstates, for our purpose of observing the energy-dependent localization spectrum, we set the initial state of the system to be an energy eigenstate of $H_s$, where we use a filling fraction of $1/4$ (that is, the total $S_z$ is $-3$). This constitutes the main result of our paper. In Sec.~\ref{llength}, we revisit the conventional setup initialized in a superposition state. We note that due to the $1/4$ filling constraint, the modified N\'eel state is now $|\downarrow\uparrow\downarrow\downarrow\downarrow\uparrow\downarrow\downarrow\downarrow\uparrow\downarrow\downarrow\rangle$.
        In both cases, the initial state of the bath is $|\downarrow\uparrow\downarrow\downarrow\downarrow\uparrow\downarrow\downarrow\downarrow\uparrow\downarrow\downarrow\rangle$. In the energy-resolved setup, we take the average over 14 choices of the initial phase $\phi$ and 11 consecutive energy levels, so that the 220 energy levels of the system are grouped into 20 data points.
        In the superposition setup, we take an average of over 342 choices of $\phi$.
        The random choices of $\phi$ are always fixed throughout the entire study so that the relative errors between the data points are lower.
	
	Our main observable is the growing entanglement entropy between the system and the bath for $t=500,510,\ldots,1000$. The time-dependent entanglement entropy is given by
	\begin{equation}
		S(t) = -\text{Tr}\left(\rho_s(t)\ln \rho_s(t)\right),~\rho_s = \text{Tr}_b \ket{\psi(t)}\bra{\psi(t)},
	\end{equation}
	where $\ket{\psi(t)}$ is the system-bath wavefunction at time $t$ and $\text{Tr}_b$ traces out the bath, leaving only the density matrix of the system.
	In the superposition case, we also calculate the spin imbalance,
	\begin{equation}
		I(t)=D\sum_{j=2}^{L_s} \langle\psi(t)|S_j^z|\psi(t)\rangle\langle\psi(0)|S_j^z|\psi(0)\rangle,
	\end{equation}
	where $D$ is chosen so that $I(0)=1$. Note that the first spin of the system is skipped to avoid any direct effect due to $U_{sb}$.

          We note that the time step $\tau=10$ and the time range $t=500$--$1000$ are most suitable for our purpose. One should not think of our setup as a Trotterized approximation of the evolution under a constant Hamiltonian, and thus $\tau$ does not need to be small but can be tuned to clearly differentiate the three behaviors of the GDP model. Indeed, a small $\tau$ both increases the computational expense and induces strong variation on $S_\text{max}$, which in turn complicates the fitting process (three fitting parameters instead of two as in the sections below).
          Therefore, the ``kicked'' setup with large $\tau$ actually help us by increasing and stabilizing $S_\text{max}$, so that the dependence of $S(t)$ on the system parameters becomes simpler.
          Lastly, we numerically verify that $\tau=5$ and $\tau=15$ give essentially the same result, so our finding is not a fine-tuned effect on $\tau$.

          Regarding the time range, the minimum time should be at least long enough such that the complicated transient dynamics is suppressed.
          We have seen that after $t\sim 500$, the entropy variation with time become smooth enough for our purpose (see Appendix~\ref{sec:raw}).
          For the upper bound of the time window, we note that for small $W$, beyond $t\sim 10^3$, $S(t)$ will be too close to $S_\text{max}$ that the dynamics is severely overwhelmed by numerical fluctuation.

	\section{Three-phase classification} \label{threephases}

	\begin{figure*}
		\includegraphics[trim=0 0 0 0, clip, width=\textwidth]{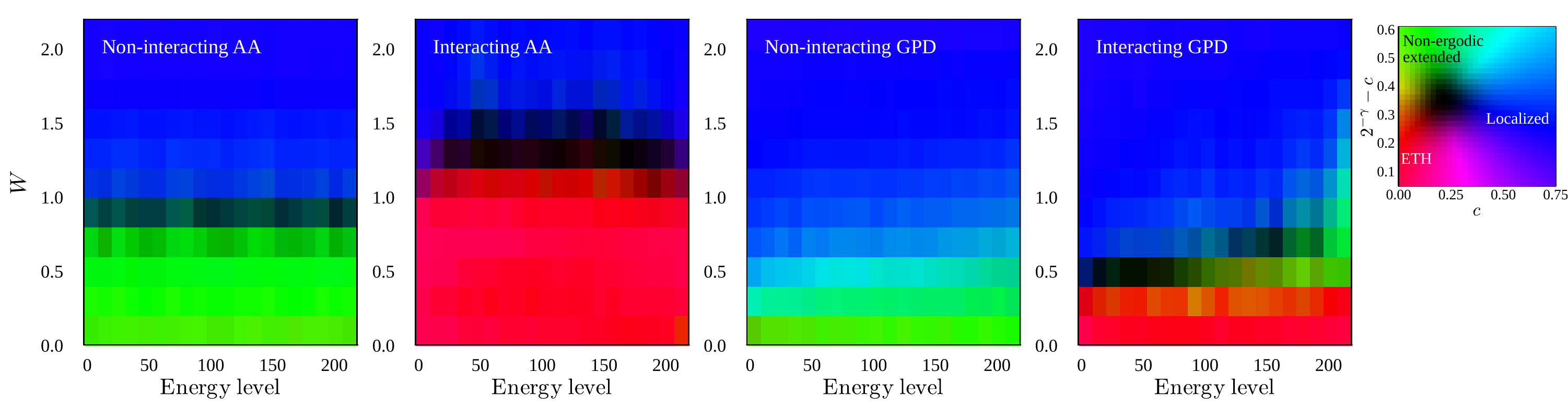}
		\caption{
                  The false-color ``phase diagrams'' obtained by mapping the $(E, W)$ parameter space to the $(c,2^{-\gamma}-c)$ feature space using the same data is the same as in Fig.~\ref{fig:circle}.
			The rightmost figure is the legend (the feature-to-color mapping), where the shaded region roughly corresponds to the dashed circles in Fig.~\ref{fig:circle}
			Note that red, green, and blue correspond to the ETH, the non-ergodic extended, and the localized behavior, respectively.
			We can see that this feature space gives the expected phase diagrams in the known parameter regime (note that the non-interacting GPD phase diagram shows an effect of the mobility edge, although not very sharp due to the small system size).
			For the interacting GPD model where the behavior is not yet fully understood, it is clear that the three behaviors coexist in the phase diagram indicated by the three colors, unlike the other three cases where only two behaviors are present (green and blue for non-interacting systems; red and blue for interacting systems), suggesting a non-ergodic metallic phase there.
		}
		\label{fig:phase}
	\end{figure*}
	
	Our classification scheme is based on the long-time dynamics of the entanglement entropy between the system and the bath, distinguishing the ETH, non-ergodic extended, and localized regimes. To extract the features from $S(t)$, we first define the quantity
	\begin{equation}
		\tilde S(t) = \frac{S_\text{max}-S(t)}{S_\text{max}}
	\end{equation}
	where $S_\text{max}\approx 5.6088$ is the empirical saturated entropy of the setup.
	Note that $\tilde S(0)=1$ and $\tilde S\to 0$ as $t\to\infty$.
        In the time range of $t=500$--$1000$, $\tilde S$ can be fitted to a power law decay (see Appendix \ref{sec:raw})
	\begin{equation}\label{poweransatz}
          \tilde S(t) \approx c\left(\frac{t}{t_0}\right)^{-\gamma},
	\end{equation}
      where we use $t_0=500$ and $c,\gamma>0$ are the fitting parameters. We find that $c$ and $\gamma$ together can classify different regimes in the $(E,W)$ parametric space. In particular, the three regimes: ETH, non-ergodic extended, and MBL reside on three corners of the $(c,2^{-\gamma}-c)$ feature space. 

           We emphasize that a simple power law is probably incapable of describing the dynamics up to an infinitely long time (equivalently the Heisenberg time $t_H\sim 70000$), but the fitting ansatz~\eqref{poweransatz} together with time window $t=500$--$1000$ is most suitable for distinguish the three different behaviors.
	
        Before describing numerical results, we provide a physical interpretation of the feature space. Our two indicators, $c$ and $2^{-\gamma}-c$, can be written as
	\begin{align}
		c &\approx \frac{\tilde S(t_0)}{\tilde S(0)}, \\
                2^{-\gamma}-c &\approx \frac{\tilde S(2t_0)}{\tilde S(t_0)}  - \frac{\tilde S(t_0)}{\tilde S(0)}.
	\end{align}
        We can see that $c$ quantifies the decay of $\tilde{S}(t)$ (or the saturation of $S(t)$) until $t_0$, characterizing the early-time thermalization rate. On the other hand, $2^{-\gamma}-c$ compares the thermalization rate at late time $t_0$--$2t_0$ with that at early time $0$--$t_0$, measuring the deceleration over time. It is easy to see that the limit $c\to 1$ corresponds to the MBL phase, and $2^{-\gamma}-c$ is small trivially in that case because both late-time and early-time thermalization rates vanish. Small $c$, on the other hand, is a signature of the extended phase. This is because early-time localization is dominated by direct resonance, which is stronger in the extended phase \cite{Long2022}. However, $c$ alone is insufficient to distinguish between ETH and non-ergodic extended phases, necessitating the deceleration measure $2^{-\gamma}-c$. Following Ref.~\cite{Crowley2020}, the avalanche only happens partially in the non-ergodic extended phase. At early time, some $l-$bits quickly hybridize with the thermal seed, leading to fast thermalization. Unlike the ETH phase, where the thermal seed can expand infinitely to fill the entire system, the non-ergodic extended phase, at a late time, comprises disconnected thermal seeds and a finite remaining fraction of $l-$bits whose couplings with the thermal regions are too weak to fuel further avalanche. As a result, the thermalization rate at late time is significantly smaller than that at early time. Therefore,  $2^{-\gamma}-c \sim 0$ (finite) indicates the ETH (non-ergodic metallic) phase through the direct estimate of the thermalization deceleration between short time and long time.
	
        At each point of $(E, W)$ parametric space, where $E$ is fixed by the quantum state initialization and $W$ is the effective disorder parameter of the Hamiltonian~\eqref{eq:pot}, we simulate the time-dependent entanglement entropy of the system and extract the two feature indicators $c$ and $2^{-\gamma}-c$ from Eq.~\eqref{poweransatz}. In Fig.~\ref{fig:circle}, we display the calculated parametric points in the feature space with the colors denoting $W$, and the direction from head (circle) to tail indicates the variation from high to low $E$. We first note that the non-interacting models completely lack the ETH phase (vanishing $c$ and $2^{-\gamma}-c$) because these systems are always trivially non-ergodic due to the absence of any interaction. In addition, the non-interacting AA model does not show any significant energy dependence (note the length of the tail), while the GPD counterpart exhibits clearly defined tails which for intermediate $W$ stretch from the localized phase at the low-energy end to the non-ergodic extended phase at the high-energy spectrum part. These observations agree well with established facts for non-interacting AA and GPD models. 
	
	In the presence of interactions, the ETH regime in the feature space becomes populated by low-$W$ points while high-$W$ points remain MBL. These are general characteristics of any interacting quasi-periodic model. However, GPD differs from AA in two essential aspects. First, the non-ergodic extended phase prominently exists in the interacting GPD model but is almost non-existent in the AA counterpart. Secondly, for AA, the energy dependence follows the generic rule: more localized in the high and low-energy tails of the spectrum and more extended in the middle (note that the heads and tails in the second panel of Fig.~\ref{fig:circle} almost draw closed loops). On the contrary, the interacting GPD spectra display a distinct unique trend with respect to energy, with the low-energy part being more localized and the high-energy part being more extended, thus showing non-ergodic behavior. These results suggest that the well-known single-particle mobility edge in the GPD model will likely survive interaction and manifest as a non-ergodic metallic phase, which is absent in a generic interacting quasi-periodic model without mobility edges.  
	
	A false-color diagram of the 2D-to-2D parameter-to-feature mapping is shown in Fig.~\ref{fig:phase}.
	The behavior of the four systems (mapped color) can be summarized as follows
	\begin{itemize}
		\item Non-interacting AA ($V=0$, $\alpha=0$)\\
		At low disorder $W\lesssim 1$, it shows the expected non-ergodic extended behavior (green), and for $W\gtrsim 1$, it shows the expected localized behavior (blue), with virtually no energy dependence, consistent with the whole spectrum being either extended or localized depending on $W<1$ or $>1$.
		\item Interacting AA ($V=1$, $\alpha=0$)\\
		At low disorder $W\lesssim 1.5$, it shows the expected ETH behavior (red); for $W\gtrsim 1.5$, it shows the localized behavior (blue), with virtually no energy dependence.
		There is a slight energy dependence in the sense that the edge part of the spectrum is slightly more localized, as expected.
Thus, interacting AA is either ETH or MBL, depending on $W$.
		\item Non-interacting GPD ($V=0$, $\alpha=-0.8$)\\
                  $W\lesssim 0.3$ shows the expected non-ergodic extended behavior for the entire spectrum (green), and for $W\gtrsim 1.1$ it shows the expected localized behavior for the entire spectrum (blue).
                  Within the intermediate disorder regime (i.e.\ for $0.3\lesssim W \lesssim 1.1$), the high energy part is more extended (more greenish via the intermediate color of cyan), as expected from the existence of a single-particle mobility edge.
		\item Interacting GPD ($V=1$, $\alpha=-0.8$)\\
		$W\lesssim 0.3$ shows the expected ETH extended behavior for the entire spectrum (red), and for $W\gtrsim 1.7$ it shows the expected localized MBL behavior for the entire spectrum (blue).
		In between, we see that the low energy part shows the localized behavior, while the high energy part shows the non-ergodic extended behavior (green).
		This supports the claim that three phases, including a non-ergodic metallic phase, exist in the interacting GPD model's $(E, W)$ parameter space.
	\end{itemize}

	\section{Localization length in intermediate and strong disorder regimes} \label{llength}

	\begin{figure*}
		\includegraphics[trim=0 0 0 0, clip, scale=0.62]{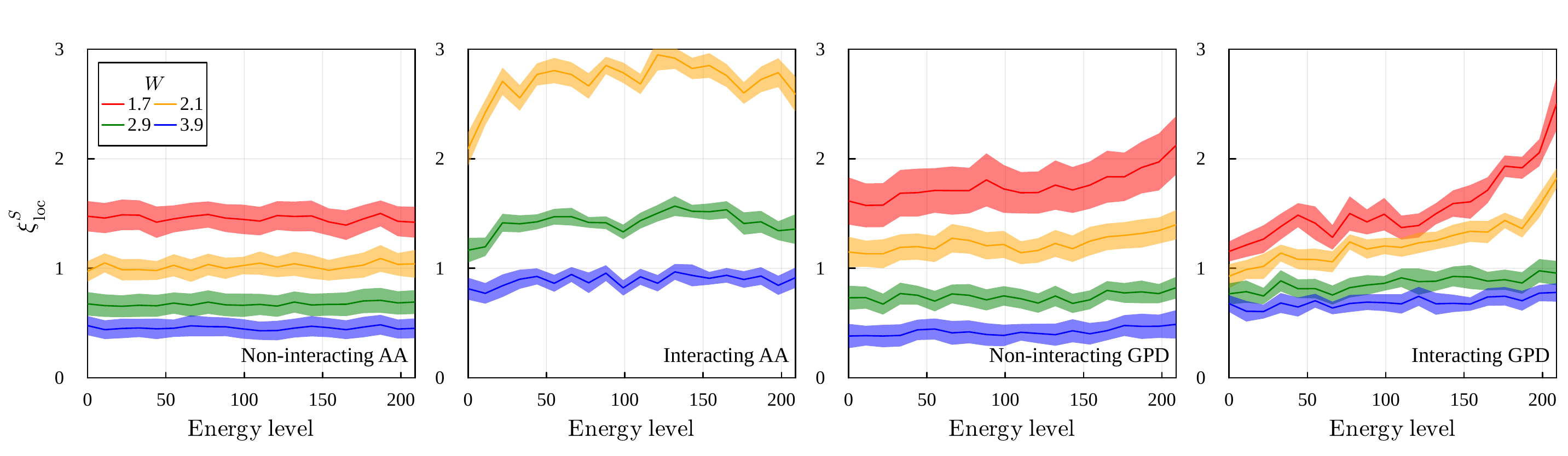}
		\caption{
			The energy-resolved localization length $\xi_\text{loc}$ extracted from the long-time behavior of $\tilde S(t)$.
			The $W=1.7$ curve of interacting AA is not shown because it is too close to the ETH phase.
			Ribbons indicate the symmetrized $68\%$ bootstrap confidence interval for the randomness of the choices of $\phi$.
		}
		\label{fig:loc}
	\end{figure*}
	
	\begin{figure}
		\includegraphics[trim=0 0 0 0, clip, scale=0.62]{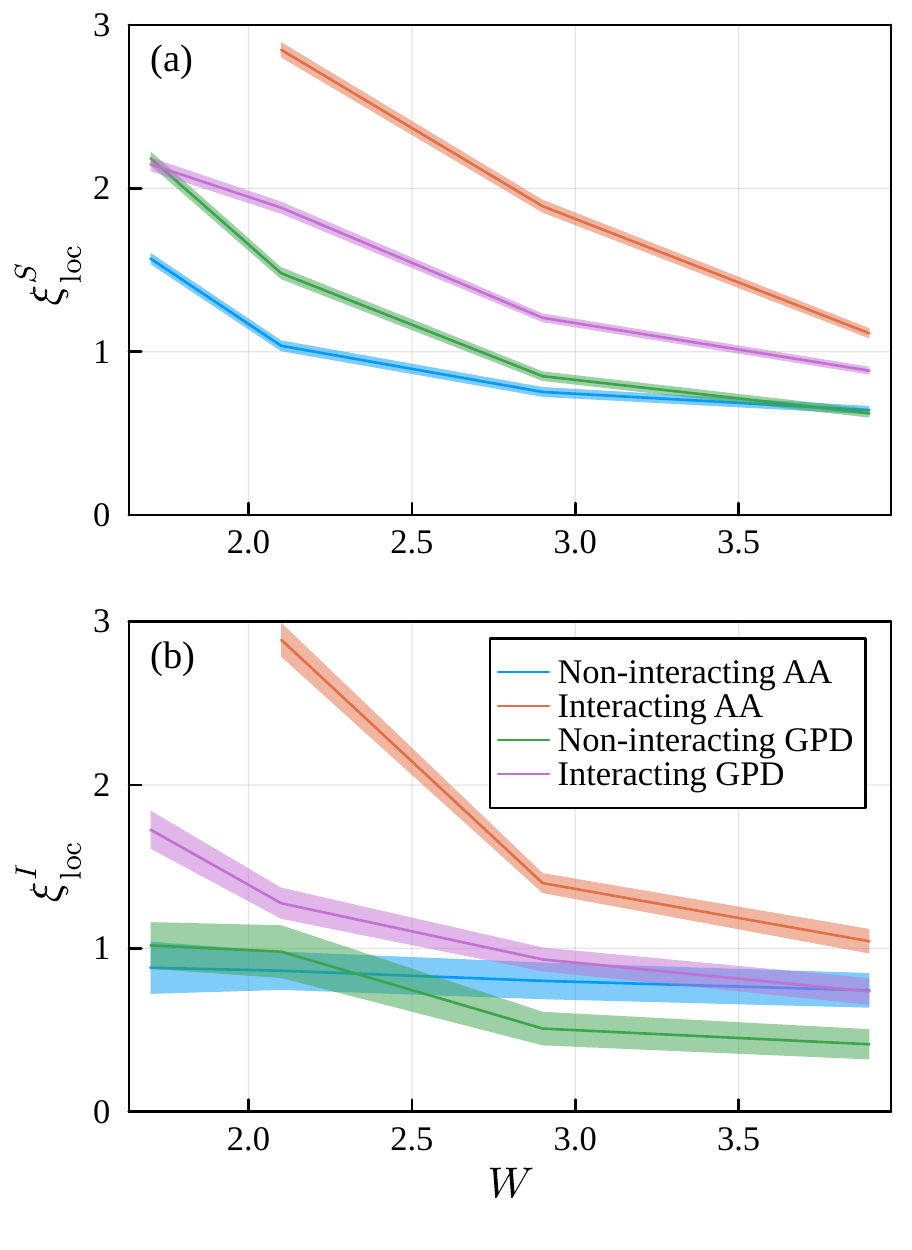}
		\caption{
			The superposition localization length $\xi_\text{loc}$ extracted from the long-time behavior of (a) $\tilde S(t)$ and (b) $\tilde I(t)$.
			Ribbons indicate the symmetrized $68\%$ bootstrap confidence interval.
		}
		\label{fig:unres}
	\end{figure}
	
        The previous section uses the power-law ansatz to classify three phases. For the MBL phase, the $\gamma$ at a late time is so low that some information is hidden, such as the $l-$bit structure where thermalization happens through long-range resonance being suppressed exponentially with distance. For this purpose, we extract the localization length in the localized phase using the long-time dynamics of the entanglement entropy and the spin imbalance.
	In the localized phase, we expect that the dynamics at the time scale $t=500$--$1000$ is dominated by the bath coupled directly with the degrees of freedom in the system (the time scale is not long enough for the avalanche effect).
	Since these degrees of freedom (DOFs) are localized, we can model the coupling strength of a DOF at a distance $j$ away from the bath as~\cite{DeRoeck2017, Thiery2018, Crowley2020}
	\begin{equation}
		g_j=g_0 e^{-\frac{j}{\xi_\text{loc}}}
	\end{equation}
	where $\xi_\text{loc}$ is the localization length of the DOF.
	The normalized decay of the part of $\tilde S$ that is due to the bath entangling with this DOF can be modeled by $f(g_j^2 t)$, where $f$ is a function satisfying $f(0)=1$, $f(+\infty)\to 0$.
	The exact form of $f$ is not essential, but it is expected that, on the $\ln t$ scale, $f$ is initially constant at $1$ and then goes down to $0$ within a scale several times smaller than $\ln t_\text{max}\approx 7$, and then stays constant at $0$.
	Now the decay of the full $\tilde S(t)$ can be modeled by
	\begin{equation}
		\tilde S(t) = \frac{1}{N} \sum_j f(g_j^2 t),
	\end{equation}
	where $N$ is the number of DOFs. Since we are averaging several energy levels and disorder realizations, the $j$ in the sum runs over a large number of essentially uniformly distributed samples within $L_s$, and $N$ is the number of such samples.
	
	Now on the $\ln t$ scale, each term in the sum goes from $1$ to $0$ fairly quickly near $\ln t \approx \ln(1/g_j^2)= 2j/\xi_\text{loc}+\text{const.}$.
	This implies that if we increase $\ln t$ from $a$ to $b$, the DOFs with $j$ within the range $a\lesssim 2j/\xi_\text{loc}+\text{const.}\lesssim b$ will decay, which is approximately $N (b-a)\xi_\text{loc} / (2L_s)$ of them in the time regime where the decaying is neither near the beginning nor the end.
	Since the decay of each DOF contributes to the change by $-1$ in the sum (although the form of $f$ smooths out the sum, the slope is not affected), this gives a constant slope of $\tilde S(t)$ in the $\ln t$ scale
	\begin{equation}\label{eq:log}
		\tilde S(t) \approx - \frac{\xi_\text{loc}^S}{2L_s} \ln t + \text{const.}
	\end{equation}
	in this time regime. (The superscript $S$ indicates that it is extracted using the entropy.)
	Note that the $\ln t$ fit does not contradict the power law fit used in the previous section, as the power $\gamma$ approaches zero as the system becomes more and more localized, essentially becoming a log behavior.
	
        In our case, we fit this slope using $\tilde S(t)$ for $t=500,510,\cdots,1000$ to extract $\xi_\text{loc}$. This ansatz is valid if the fast hybridizing modes are thermalized before $t=500$ and the slow modes survive after $t=1000$. The result of the energy-resolved case is shown in Fig.~\ref{fig:loc} and the superposition case in Fig.~\ref{fig:unres}a.
	Note that the result is consistent with the many-body inverse participation ratio given in Appendix~\ref{sec:MIPR}.
	
        The argument above also applies for the spin imbalance $I(t)$.
        However, due to the local dynamics of the spins, we also need to extract the value $I_0$ that corresponds to the beginning of the decay numerically, which is not necessarily close to 1. We use the average of $I(t)$, $t=10,20\ldots,50$ for $I_0$.
	Also, due to the $1/4$ filling and the skipping of the first spin, the equilibrium value of $I(t)$ is $I_\text{eq}\approx 0.2273$ instead of 0.
	Now we have
	\begin{equation}\label{eq:logS}
		\tilde I(t) \equiv \frac{I(t)-I_\text{eq}}{I_0-I_\text{eq}} \approx - \frac{\xi_\text{loc}^I}{2(L_s-1)} \ln t + \text{const.}
	\end{equation}
	in the same time regime as Eq.~\ref{eq:log}.
	We do the same fit as the case of $\tilde S$, and the result is shown in Fig.~\ref{fig:unres}b.
	
	From Fig.~\ref{fig:unres}, it is clear that, although AA is more stable than GPD in the non-interacting case, the interaction destabilizes AA dramatically while only making GPD slightly more unstable, resulting in GPD being more stable than AA in the interacting case.
	From Fig.~\ref{fig:loc}, it is also clear that non-interacting AA is more stable than GPD due to the high energy part of the GPD spectrum, which is more extended.
	The interaction, on the other hand, stabilizes the lower part of the spectrum of GPD slightly while destabilizing the upper part.

	\section{Conclusion} \label{conclude}
	By coupling to a thermal bath, we observe three distinct regimes of the interacting GPD model: MBL, ETH, and non-ergodic extended. While the former two phases are universal for most models with single-particle localization, the third one is more elusive, not even universal to quasi-periodic models with single-particle mobility edges. This suggests a peculiar richness of the GPD model and invites deeper examination.
	
	Our probing method also adds a new perspective to the standard MBL analysis, which can be generalized to other theoretical and experimental studies. The system in our work is inherently open due to an explicit bath coupling and remarkably produces qualitatively similar characteristics as the closed system. Our bath coupling probe, thus, can serve as a natural bridge between the conventional closed-system MBL and the newly emerging open-system or non-Hermittian MBL \cite{Hamazaki2019,Sascha2021,OBrien2023}.

        We mention that our results are based on finite size and finite time exact simulations, and therefore, we cannot comment decisively on the existence or not of the novel interacting nonergodic extended phase in the long time thermodynamic limit, but this limitation is no more severe in our work than it is for the basic question of the existence or not of the MBL itself in the thermodynamic limit.  In fact, we believe that the non-ergodic extended phase should manifest itself in interacting atomic systems, which are constructed to mimic the interacting GPD phase, through measurements and analysis of imbalance dynamics~\cite{An2021}.  We urge such experiments in through the analog simulation of the GPD model, given the unique unusual nature of the interacting nonergodic extended phase as analyzed in the current work.

          Although the ``kicked'' coupling setup with large $\tau$ helps us simplify the behavior of $S(t)$ by increasing and stabilizing $S_\text{max}$, one may wonder whether the non-ergodic extended phase might be originated from this setup rather than from the system itself.
Nevertheless, we do not expect it is the case. First, there have been other previous evidence that such a regime may exist in this model~\cite{Li2015a,Li2016,Hsu2018_PRL,YiTing_ML}. Second, our probe also correctly shows that there is no such regime in the interacting AA model. Therefore, the non-ergodic extended phase is likely from the GDP Hamiltonian rather than the kicked setup.
          This point may be confirmed further in the future by applying this setup to other models with different kicked setup parameters.

	\section*{Acknowledgement} The authors thank Yi-Ting Hsu for useful discussions.
	This work is supported by the Laboratory for Physical Sciences. The authors acknowledge the University of Maryland supercomputing resources (\href{https://hpcc.umd.edu}{https://hpcc.umd.edu}) made available for conducting the research reported in this paper.
	
	\appendix
	\section{Inverse participation ratio}\label{sec:MIPR}
	
	\begin{figure*}
		\includegraphics[trim=0 0 0 0, clip, scale=0.62]{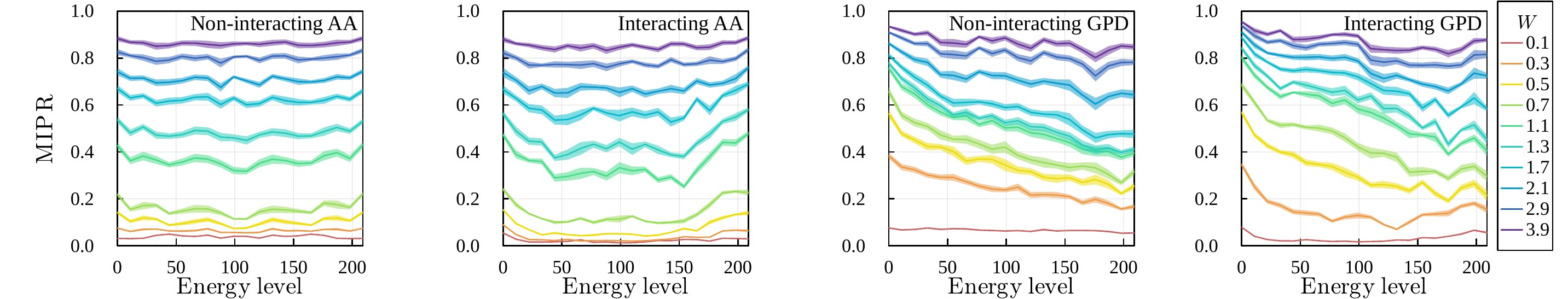}
		\caption{
			The MIPR of the energy eigenstates.
			Ribbons indicate the symmetrized $68\%$ bootstrap confidence interval for the randomness of the choices of $\phi$.
		}
		\label{fig:MIPR}
	\end{figure*}
	
	We calculate the many-body inverse participation ratio (MIPR) of the four systems we study, defined by
	\begin{equation}
		\text{MIPR}=\frac{1}{1-\nu}\left(\frac{1}{L\nu}\sum_{j=1}^L \langle n_j \rangle^2 - \nu\right),
	\end{equation}
	where $\nu=1/4$ is the filling fraction and $n_j=S_j^z+1/2$ is the particle number operator.
	The result is shown in Fig.~\ref{fig:MIPR}.
	Note that the value approaches $0$ in the extended phase and $1$ in the localized phase. The energy dependence is consistent with the results in the main text, with a monotonic variation only showing up in GPD.
	
	\section{Selective raw data of $\tilde S(t)$ and $I(t)$}\label{sec:raw}
	
	\begin{figure*}
		\includegraphics[trim=0 0 0 0, clip, scale=0.62]{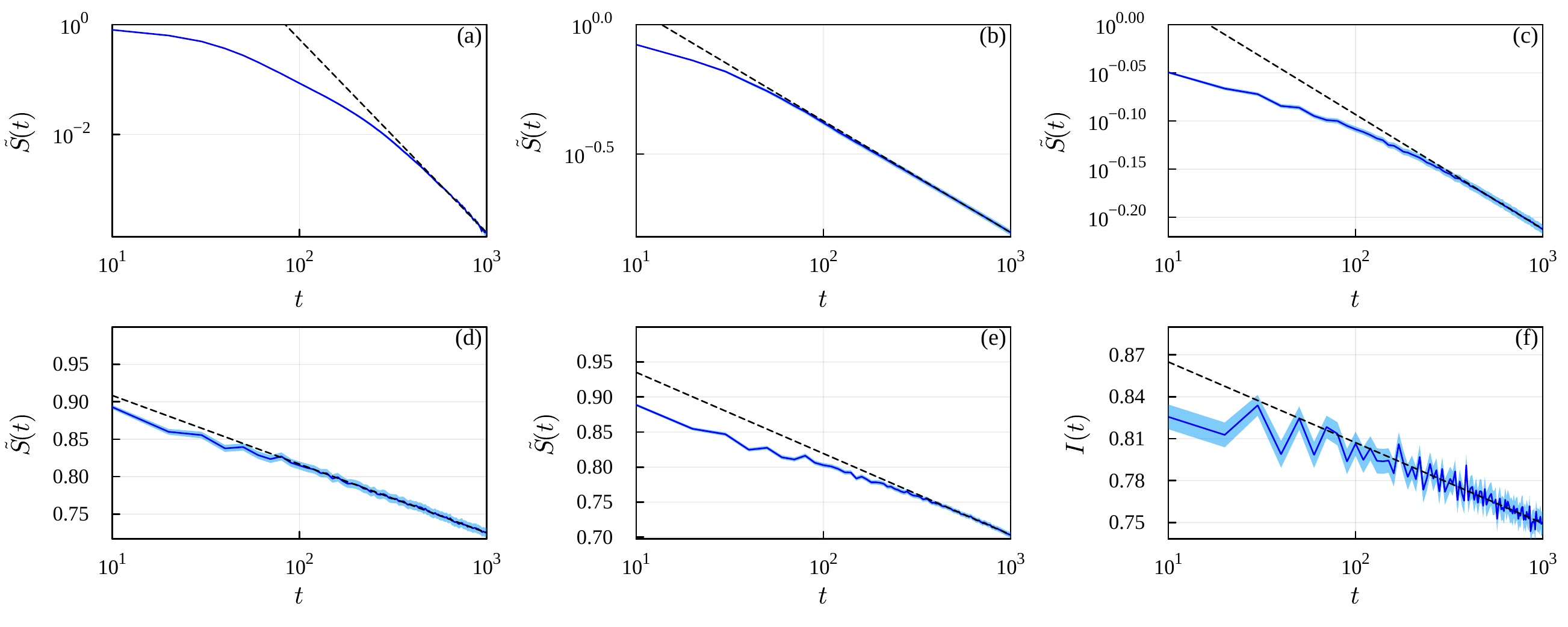}
		\caption{
			Some raw data of $\tilde S(t)$ and $I(t)$ for the interacting GPD model.
			(a-c) The power law fitting for the energy level group 210--220 for (a) $W=0.1$ (ETH), (b) $W=0.7$ (non-ergodic extended), and (c) $W=2.1$ (localized).
			(d) The log fitting for the same energy level group for $W=2.9$.
			(e--f) The superposition case log fittings for $W=2.9$.
                        Ribbons indicate the standard errors of the mean. Dashed lines indicate the best fit for $t=500$--$1000$.
		}
		\label{fig:raw}
	\end{figure*}
	
	Here we present some raw data of the entropy saturation $\tilde S(t)$ and spin imbalance $I(t)$ in Fig.~\ref{fig:raw}, where the fitting lines are used to extract the data presented in the main text ($I(t)$ requires an additional scaling indicated in Eq.~\ref{eq:logS}). We only present representative cases here; 
        the cases of other parameters are qualitatively similar to these figures. We can see that the spin imbalance is much more noisy than the entanglement entropy. (Note that Fig.~\ref{fig:raw}e--f are averaged over 342 choices of $\phi$ while a--d are only over 14.)

	\bibliographystyle{apsrev4-2}
	\bibliography{references}
	
\end{document}